\documentstyle[12pt]{article}
\begin{document}
\title{{\bf The Height of a Giraffe}
\thanks{Alberta-Thy-11-07, arXiv:0708.0573}}
\author{
Don N. Page
\thanks{Internet address:
don@phys.ualberta.ca}
\\
Institute for Theoretical Physics\\
Department of Physics, University of Alberta\\
Room 238 CEB, 11322 -- 89 Avenue\\
Edmonton, Alberta, Canada T6G 2G7
}
\date{(2007 August 3)}

\maketitle
\large
\begin{abstract}
\baselineskip 18 pt

A minor modification of the arguments of Press and Lightman leads to
an estimate of the height of the tallest running, breathing organism
on a habitable planet as the Bohr radius multiplied by the
three-tenths power of the ratio of the electrical to gravitational
forces between two protons (rather than the one-quarter power that
Press got for the largest animal that would not break in falling
over, after making an assumption of unreasonable brittleness).  My
new estimate gives a height of about 3.6 meters rather than Press's
original estimate of about 2.6 cm.  It also implies that the number
of atoms in the tallest runner is very roughly of the order of the
nine-tenths power of the ratio of the electrical to gravitational
forces between two protons, which is about $3\times 10^{32}$.

\end{abstract}
\normalsize

\baselineskip 14 pt

\newpage

\section*{Introduction}

Press \cite{Press} and Press and Lightman \cite{PL} have given
estimates of our size as being of the order of $(e/m_p)^{0.5}a_0$,
which is about 0.0558 meters or nearly 6 cm, using Planck units
$\hbar = c = G = 4\pi\epsilon_0 = k = 1$, with $k$ being Boltzmann's
constant.  (Press \cite{Press} actually included some numerical
factors that reduced his estimate to about 2.6 cm.)  Here I give a
slightly modified argument that gives a size estimate of the order
of $(e/m_p)^{0.6}a_0$, which leads to a more accurate value of 3.56
meters.

Press \cite{Press} uses three ``requirements'':  ``(i) We are made
of complicated molecules; (ii) we breathe an evolved planetary
atmosphere; (iii) we are about as big as we can be without
breaking.''  In more detail, Press states the following:

``Let us assume only that man [the term used generically to include
male as well as female] satisfies three properties:  (i) he is
made of complicated molecules; (ii) he requires an atmosphere which
is not (primordial, cosmological) hydrogen and helium; and (iii) he
is as large as possible, to carry his huge brain, but he is liable to
stumble and fall; and in so doing he should not break.  These three
properties do not differentiate between a man and, say, an elephant
of size $L_E$; however $L_E \approx L_H$ to the accuracy of our
calculation, and we should not expect to distinguish elephants from
men by dimensional arguments.''

Here I shall accept Press's assumptions (i) and (ii) but instead of
using his (iii), I shall use the assumption that the organism in
question is the tallest organism that can run without overheating. 
Strictly speaking, this would better give the height of a giraffe
rather than that of a man, and coincidentally the ignored numerical
factors conspire to make that true, but as Press mentioned in
comparing a man and an elephant, my estimate for the height of a
giraffe is also a rough estimate of our height as another species of
the largest running animals on earth.

In this argument, it is convenient to define three dimensionless
quantities, which I shall give here first in Planck units and then
in in conventional units (if the expressions differ):  (a) the
fine-structure constant,
\begin{equation}
\alpha \equiv e^2 \equiv \frac{e^2}{4\pi\epsilon_0 \hbar c}
\approx \frac{1}{137.036\,00} \approx 0.007\,297\,352\,6,
\label{eq:1}
\end{equation}
the ratio of the electron mass to the proton mass,
\begin{equation}
\beta \equiv \frac{m_e}{m_p}
\approx \frac{1}{1\,836.152\,763} \approx 0.000\,544\,617\,021\,6,
\label{eq:2}
\end{equation}
and the ratio of the electrical repulsion to the gravitational
attraction between two protons,
\begin{equation}
\gamma \equiv \frac{e^2}{m_p^2}
\equiv \frac{e^2}{4\pi\epsilon_0 G m_p^2}
\approx 1.236\times 10^{36}.
\label{eq:3}
\end{equation}

Then, for example, in terms of the Planck length $L_P \equiv
\sqrt{\hbar G/c^3} \approx 1.616\times 10^{-35}$ m and the Planck
mass $M_P \equiv \sqrt{\hbar c/G} \approx 2.176\times 10^{-8}$ kg
$\approx 1.311\times 10^{19}$ amu, which I am setting
to unity when I use Planck units, one can write the Bohr radius as
\begin{equation}
a_0 \equiv m_e^{-1} e^{-2} = \alpha^{-3/2}\beta^{-1}\gamma^{1/2}L_P
\approx 0.529\,177\,21\times 10^{-10}\, {\mathrm m}.
\label{eq:4}
\end{equation}

One can also define a crude estimate for a typical stellar mass
\cite{PL} as
\begin{equation}
M_s \equiv m_p^{-2} = \alpha^{-1}\gamma M_P
\approx 3.685\times 10^{30} {\mathrm kg} \approx 1.853\, M_\odot,
\label{eq:5}
\end{equation}
where $M_\odot \approx 1.988\times 10^{30}$ kg is the mass of the
sun.  If the mass $M_s$ were a black hole, the corresponding length
scale (half the Schwarzschild radius) would be
\begin{equation}
L_s \equiv m_p^{-2} = \alpha^{-1}\gamma L_P
\approx 2\,737\, {\mathrm m}.
\label{eq:6}
\end{equation}
Below it is convenient to write estimates of the mass $M$ and radius
$R$ of a habitable planet in terms of $M_s$ and $L_s$ and the
dimensionless ratios $\alpha$, $\beta$, and $\gamma$, and to write
animal sizes in terms of $a_0$ and these same dimensionless ratios.

\section{Press and Press-Lightman Estimates}

Requirement (i) in Press \cite{Press}, that we are made of
complicated molecules, leads to the requirement that the
environmental temperature $T$ have an energy equivalent $kT$ that is
less than the binding energy of those molecules.  On the other hand,
the mobility assumed by (iii) implies that $kT$ is not enormously
less than the binding energies, or else the internal energy
processes would occur exponentially slowly.  These requirements lead
to an environmental temperature $T$ whose equivalent energy $kT$ is
a small fraction, say $\epsilon$, of the Rydberg energy,
\begin{equation}
{\mathrm Ry} \equiv \frac{1}{2}m_e e^4 \sim m_e \alpha^2
= \alpha^{5/2}\beta\gamma^{-1/2} M_P c^2,
\label{eq:7}
\end{equation}
dropping the factor of $\frac{1}{2}$ in the final two expressions. 
Press \cite{Press} says that a reasonable value for the small
fraction $\epsilon$ is 0.003; Press and Lightman \cite{PL} take the
small fraction to be $\varepsilon(m_e/m_p)^{1/2}$ with $\varepsilon
\sim 0.1$, which would give $\epsilon = \varepsilon \beta^{1/2}
\approx 0.0023$.

Actually, Press and Lightman use the same $\epsilon$ symbol as Press
does, even though they are different quantities, but here I shall
distinguish the Press-Lightman one by writing it as $\varepsilon$. 
For simplicity in the analysis below, I shall drop numerical factors
of the rough order of unity and set the Press-Lightman $\varepsilon$
to unity to make the Press $\epsilon = \beta^{1/2} = (m_e/m_p)^{1/2}
\approx 0.0233$.  The fact that this is almost 8 times larger than
the value $\epsilon \sim 0.003$ advocated in Press \cite{Press} and
is 10 times larger than the value advocated in Press and Lightman
\cite{PL} is often partially compensated by my neglect of other
numerical factors, such as the ratio of $a_0^3$ to the space taken
up by an atom, so when I neglect these other factors, I find that it
usually doesn't help much, if any, to include the $\varepsilon$
factor.

At the end of the Press-Lightman presentation \cite{PL}, it is
recorded that Rudolf Peierls objected to
$\varepsilon(m_e/m_p)^{1/2}\,{\mathrm Ry}$ as an estimate of
molecular binding energies, asking whether
$(m_e/m_p)^{1/2}\,{\mathrm Ry}$ instead represents ``an estimate of
the zero-point energy of vibration,'' which ``would seem a
considerable underestimate, since the vibrational energy usually
amounts to many vibrational quanta.''  Press concurred, ``Yes, the
factor of $(m_e/m_p)^{\frac{1}{2}}$ does strictly give the
characteristic energy-level {\it spacing} of molecules, rather than
their full binding energy.  Numerically, however, it does also give
the correct (rough) factor by which molecular bindings are smaller
than typical atomic ones.  One may wish to consider the factor a
{\it mnemonic} for this dimensionless ratio of binding energies
(which derives from a theory of chemistry) rather than an accurate
physical `theory' of that ratio.''

In our part of the universe, where $\beta^{1/2} = (m_e/m_p)^{1/2}
\approx 0.0233$, it is thus numerically okay to use this as a rough
estimate for the ``factor by which molecular bindings are smaller
than typical atomic ones,'' and in this paper I shall do that. 
However, for other possible parts of our universe or multiverse that
have electrons and protons but with potentially greatly different
mass ratios, it may no longer be even roughly valid to use
$(m_e/m_p)^{1/2}$ for the ratio of biological molecular binding
energies to the Rydberg in that part of the universe.  Therefore,
quantities below that depend upon $\beta \equiv m_e/m_p \approx
1/1\,836.152\,673$ should be interpreted cautiously for other parts
of the universe where this small ratio $\beta$ is significantly
different.  However, in my estimate below for the height of the
largest running organism, $\beta$ enters with only the one-twentieth
power, so one can effectively drop the $\beta$ dependence (assuming
that it is not too many orders of magnitude away from unity), as I
do to get the simple formula for the giraffe height as a Bohr radius
multiplied by the three-tenths power of the ratio of the electrical
to gravitational forces between protons.

If one does set $kT = \beta^{1/2} m_e e^4 = \alpha^2 (m_e/m_p)^{1/2}
m_e c^2$, where the last expression reverts from Planck units to
ordinary units by inserting $c^2$, one gets that $T \approx 7\,369
K$, which of course is far hotter than the surface of the earth. 
One would get a much better result by inserting the factor of $1/2$
for the Rydberg and a factor of $\varepsilon = 0.1$, which would
give $T \approx 368 K \approx 95 C$, which would still be unbearably
hot for humans (nearly boiling) but which would be within 30\% of a
typical earth surface value.  However, since it is both {\it ad hoc}
and difficult to give good estimates of all the numerical factors
such as $\varepsilon$, here I shall ignore all of them and just
proceed with
\begin{equation}
kT \sim m_e^{3/2} m_p^{-1/2} e^4 =
\alpha^2\beta^{1/2}m_e c^2 \approx 0.636\, {\mathrm eV},
\label{eq:8}
\end{equation}
where for the penultimate expression on the right I have inserted
the factor of $c^2$ that is unity in Planck units in order to have
the correct expression in ordinary units, with the small
dimensionless quantities $\alpha$ being the fine-structure constant
and $\beta$ being the ratio of the electron mass to the proton mass.

Requirement (ii) in Press \cite{Press}, that we breathe an evolved
planetary atmosphere, leads to a habitable planet of radius $R$ and
mass $M \sim (m_p/a_0^3)R^3$ such that the magnitude of the
gravitational binding energy of hydrogen, $GMm_p/R$, is of the order
of $\epsilon kT \sim \beta^{1/2}{\mathrm Ry} \sim (m_e/m_p)^{1/2}m_e
e^4$, so that hydrogen and helium can escape from the earth, but not
most of the heavier gases.  For simplicity I am dropping numerical
factors of the order of unity for the volume and density of the
earth as well as the Press-Lightman factor $\varepsilon$ and the
factor of $1/2$ in the Rydberg.  This then leads to a habitable
planetary radius
\begin{equation}
R \sim m_e^{-3/4} m_p^{-5/4} e^{-1}
= \alpha^{-3/2}\beta^{-3/4}\gamma L_P
= \alpha^{-1/2}\beta^{-3/4} L_s
\approx 8\,986 {\mathrm km} \approx 1.409 R_\oplus,
\label{eq:9}
\end{equation}
\begin{equation}
M \sim m_e^{3/4} m_p^{-11/4} e^{3}
= \alpha^{1/2}\beta^{3/4}\gamma M_P
= \alpha^{3/2}\beta^{3/4}M_s
\approx 8.190\times 10^{24} {\mathrm kg} \approx 1.371 M_\oplus,
\label{eq:10}
\end{equation}
where $R_\oplus \approx 6\,378.140$ km is the equatorial radius of
the earth and $M_\oplus \approx 5.972\times 10^{24}$ kg is the mass
of the earth.

Thus although ignoring the factors of $1/2$ and $\varepsilon$ made
the estimated temperature $T$ come out about 25 times a typical
earth surface temperature, for the mass and radius of a habitable
planet these factors are mostly canceled by other numerical factors
that I am also ignoring, so both $M$ and $R$ are within 37--41\% of
the values for earth.  It is amusing that the resulting crude
estimate for the orbital speed of a satellite skimming the planet
comes out to be very accurate,
\begin{equation}
v_{\mathrm satellite} \sim \alpha\beta^{3/4}
\approx 0.2602\times 10^{-6} c \approx 7.799\, {\mathrm km/s},
\label{eq:11}
\end{equation}
which is just 1.34\% smaller than the actual value of
$\sqrt{GM_\oplus/R_\oplus} \approx 0.2637\times 10^{-6} c \approx
7.905$ km/s for the earth.

For the properties of life on a planet, the main parameters of
importance are the temperature $T$ estimated above and the surface
gravity that may be estimated as
\begin{eqnarray}
g = \frac{GM}{R^2} \sim m_e^{9/4} m_p^{-1/4} e^{5}
&=& \beta^{1/4} m_e^2 e^5 = \alpha^{5/2} \beta^{1/4} m_e^2
= \alpha^{7/2}\beta^{9/4}\gamma^{-1} c^2 L_P^{-1} \nonumber \\
&=& \alpha^{5/2}\beta^{9/4} c^2 L_s^{-1}
\approx 6.769\, {\mathrm m/s^2} \approx 0.6903\, g_\oplus,
\label{eq:12}
\end{eqnarray}
where $g_\oplus = 9.80665$ m/s$^2$ is the standard gravitational
acceleration on earth.

Having dropped the $\varepsilon$ factor of Press and Lightman
\cite{PL}, my $\beta$ factor is essentially the factor $\epsilon^2$
of Press \cite{Press}, so one might also write this approximation as
$g \sim \epsilon^{1/2} e^5 m_e^2$, which Carter \cite{Carter} did. 
He then noted that it is interesting that if one takes $\epsilon$
(and not $\varepsilon$) to be independent of $\beta$ (as it might be
in reality, with $\epsilon \sim \beta^{1/2}$ being only a mnemonic
formula that works in our part of the universe and not universally),
then the approximation for the acceleration of gravity $g$ on a
habitable planet depends only on the properties of the electron, and
not upon the mass of the proton.  However, for agreement within 31\%
of the acceleration of gravity on earth, it does help to use
$\beta^{1/2}$ for $\epsilon$, since dropping the $\beta^{1/4}$ or
$\epsilon^{1/2}$ factor altogether gives the more memorable but more
crude estimate $g \sim e^5 m_e^2 \approx 44.31\, $m/s$^2$ $\approx
4.519\, g_\oplus$ that is nearly 5 times too large numerically.

Requirement (iii) in Press \cite{Press}, that we are about as big as
we can be without breaking when falling, led him to estimate that
the energy released by a human of size $L_H$ and mass $M_H \sim
(m_p/a_0^3)L_H^3$, $E \sim M_H g L_H$, be of the order of the energy
needed to break the human by disrupting a two-dimensional surface
containing of the order of $(M_H/m_p)^{2/3}$ atoms, which is of the
order of $(L_H/a_0)^2$.  Taking the energy needed per atom to be
$\epsilon {\mathrm Ry}$, this then is equivalent (modulo numerical
factors of order unity that I am dropping) to saying that the weight
of a proton, $m_p g$, be comparable to $\epsilon$ times the
electrical force between two protons separated by the distance
$L_H$, that is, to $\epsilon e^2/L_H^2$ in units with
$4\pi\epsilon_0=1$.

Dropping the $\epsilon$ factor for the moment for simplicity, this
last description of Press's criterion (not explicitly stated as such
in Press \cite{Press} or in Press and Lightman \cite{PL}) may be
explained as follows:  If $N_c = L_H/a_0$ represents the number of
atoms in a column one atom thick (each atom of mass approximated to
be $m_p$, since we are ignoring factors of both the atomic mass
number $A$ and the nuclear charge number $Z$), then the Press
criterion is that the energy released by this column of atoms
falling a distance $L_H$, namely roughly $N_c m_p g L_H$, is
comparable to the energy needed to break a chemical bond, which when
one sets $\epsilon = 1$ is roughly a Rydberg or roughly the
electrostatic potential energy of two protons separated by a
distance $a_0$, namely $e^2/a_0$.  Because of the $1/r$ falloff of
the electrostatic potential, this is the same potential energy as
that between one proton and a collection of $N_c$ protons (ignoring
the mutual potential energy between those $N_c$ protons) at a distance
$N_c$ times greater than $a_0$, which is $L_H$.  That energy in turn
is the force between the one proton and the $N_c$ protons at
separation $L_H$, multiplied by this separation distance $L_H$,
namely $N_c (e^2/L_H)^2 L_H$.  Equating this to the energy of fall,
roughly $N_c m_p g L_H$ by the Press criterion with $\epsilon = 1$,
gives $m_p g \sim e^2/L_H^2$, which is the condition that the proton
weight be comparable to the electrostatic force between two protons
separated by $L_H$.  When one reinserts the factor of $\epsilon$ or
$\beta^{1/2}$, one gets that Press's criterion is that the proton
weight should be about $\epsilon$ or $\beta^{1/2}$ times smaller
than the electric force between two protons separated by a distance
$L_H$ given by the size of the organism.

If I now use the Press-Lightman \cite{PL}
approximation (when their $\varepsilon$ factor is dropped) that the
Press $\epsilon \sim \beta^{1/2}$, then I get that Press's estimate
for the height of man is
\begin{equation}
L_H \sim \beta^{1/8} m_e^{-1} m_p^{-1/2} e^{-3/2}
= m_e^{-7/8} m_p^{-5/8} e^{-3/2}
= \beta^{1/8} \gamma^{1/4} a_0
\approx 0.02181\, {\mathrm m}.
\label{eq:13}
\end{equation}

If one replaces $\beta$ by Press's $\epsilon^2$, then Press's
expression \cite{Press} agrees with the first expression on the
right hand side above, except that Press has a numerical factor of 2
that I have dropped.  He writes the numerical value of $L_H$ as
$2.6(\epsilon/0.003)^{1/4}$ cm.

Press \cite{Press} recognizes that his estimated value is about 100
times smaller than the observed value and notes that he has
underestimated man's breaking strength by a factor of about
10\,000--100\,000 by assuming ``implicitly that man was
`brittle,' i.e., that the energy of the fall would be concentrated
as stress along his weakest fault plane. \ldots  Probably the reason
for this excess strength [over the estimate above] is that man's
molecular structure is polymeric rather than amorphous, so that
stresses are distributed over a rather larger volume than that of a
single monatomic fault plane.''

Press and Lightman \cite{PL} note that this size estimate that gives
roughly 3 cm is the same as ``the maximum size of water drops
dripping off a ceiling.''  Barrow and Tipler \cite{BT} write, ``The
size estimates given by Press are a better estimate of the size of a
creature able to support itself against gravity by the surface
tension of water which is some fraction of the intermolecular
binding energy, say $\epsilon\alpha^2 m_e$ per unit area, and
Press's size limits, $\sim 1$ cm, more realistically correspond to
the maximum dimension of pond-skaters rather than people.''

\section{Revised Estimates}

Brandon Carter \cite{Carter} has recently suggested that instead of
Press's third requirement, one should think of a basic biological
velocity $\tilde{v}$ such that $\tilde{m}\tilde{v}^2 \sim kT \sim
\epsilon e^4 m_e$ with $\tilde{m} \sim \tilde{\epsilon}^{-1}m_p$
being ``a mass scale characterising relevant large biochemical
molecules such as proteins.''  Carter suggests that his new small
numerical factor $\tilde{\epsilon}$ ``might tentatively be taken to
be given by $\tilde{\epsilon} \approx 10^{-3}$.''  With his
formulation of $g \approx \epsilon^{1/2} e^5 m_e^2$, Carter suggests
that the maximum height difference $\tilde{\ell}$ between different
parts of the organism will be given by a formula that implies
\begin{equation}
\tilde{\ell} \sim \tilde{v}^2/g
\sim \epsilon^{1/2} \tilde{\epsilon} e^{-1} m_e^{-1} m_p^{-1}
= \epsilon^{1/2} \tilde{\epsilon} \gamma^{1/2} a_0
\approx 3\,200\, {\mathrm m},
\label{eq:14}
\end{equation}
where I have used Press's value $\epsilon \sim 3\times 10^{-3}$ and
Carter's value $\tilde{\epsilon} \sim 10^{-3}$ to make the numerical
evaluation given at the end.

This value is over 100\,000 times the estimate of Press for the
height of man and is unfortunately too large by a factor of about
1\,000.  If one said that $\epsilon \sim \beta^{1/2}$ and that
$\tilde{\epsilon} \sim \beta$, just as I wrote Press's estimate as
$L_H \sim \beta^{1/8} \gamma^{1/4} a_0 \approx 0.022$ m in terms of
just $\beta$, $\gamma$, and the Bohr radius $a_0$, so I could do the
same to express Carter's estimate as $\tilde{\ell} \sim \beta^{5/4}
\gamma^{1/2} a_0 \approx 4\,900$ m.

It is helpful that Carter has found an argument that gives a higher
power of $\gamma$, the huge ratio ($\sim 10^{36}$) between the
electrical and gravitational forces between two protons, but
unfortunately his argument seems to lead to too great a power. 
However, it motivated me to think of the following third argument
that incorporates some of the reasoning of Press and Lightman
\cite{PL} of how hard we can work and how fast we can run:

Press and Lightman \cite{PL} argue that the power of a large animal,
say the horsepower, is ``limited by cooling through the animal's
surface area, and that resting metabolism is scaled to keep the
resting animal tolerably warm.''  They give the estimate,
\begin{equation}
({\mathrm horsepower}) \sim \Delta T \times
({\mathrm conductivity})\times
({\mathrm area})/({\mathrm skin\ depth}).
\label{eq:15}
\end{equation}

They go on to note, ``If we had no knowledge of the observed
parameters, we could use $\Delta T \approx T$, area of order $h^2$,
skin depth of order $h$ (where $h$ is given by [Press's estimate]),
and conductivity as given by (10),'' which I can write first in
their notation and then in Planck units (after dropping the factor
of $1/2$ in the definition of the Rydberg) as
\begin{equation}
({\mathrm conductivity})
\sim ({\mathrm Ry}/a_0\hbar)(m_e/m_p)^{\frac{1}{2}}k
\sim \alpha^3 \beta^{1/2} m_e^2 = \alpha^4 \beta^{5/2} \gamma^{-1}.
\label{eq:16}
\end{equation}

Then using $T \sim \alpha^2\beta^{1/2}m_e$ in Planck units from Eq.\
(\ref{eq:8}) gives the power in Planck units as
\begin{equation}
P \sim ({\mathrm conductivity}) T h \sim \alpha^5 \beta m_e^3 h,
\label{eq:17}
\end{equation}
in terms of the size $h$ of the organism.  However, here I shall not
follow Press and Lightman \cite{PL} in using Press's estimate
\cite{Press} for the height of man, $h \sim L_H$.

Press and Lightman go on to ask, ``How fast can we run?'':  ``We run
in an extremely dissipative fashion.  To run at velocity $v$, we
must renew practically our whole kinetic energy every stride, that
is, every motion through our body length $h$.  The power needed to
run at velocity $v$ is therefore of order $mv^3/h$.''  Therefore, we
shall set
\begin{equation}
P \sim mv^3/h.
\label{eq:18}
\end{equation}
Here $m$ is the mass of the organism, which at the density of roughly
$m_p$ per cubic Bohr radius, just as assumed above for the earth,
gives for an organism of volume $h^3$
\begin{equation}
m \sim (m_p/a_0^3)h^3 \sim \alpha^3 m_e^3 m_p h^3.
\label{eq:19}
\end{equation}

Now I shall follow Carter's \cite{Carter} suggestion of $\tilde{\ell}
\approx \tilde{v}^2/g$ and set $h \sim v^2/g$, though with slightly
different motivation.  Carter argues, ``Assuming that such a
velocity $\tilde{v}$ characterizes the relevant energies, pressures,
and tensions (as involved for example in the pumping of blood) in an
organism, it will provide an upper limit $g\ell \stackrel{<}{\sim}
\tilde{v}^2$ on the supportable value of the gravitational energy
per unit mass associated with a height difference $\ell$ between
different parts of the body of the organism.  This suggests that a
biological land (though not necessarily sea) organism will be able
to have a maximum size $\tilde{\ell}$ and a corresponding biological
clock timescale $\tilde{\tau}$ given by
\begin{equation}
\tilde{\tau} \approx \frac{\tilde{\ell}}{\tilde{v}}
\approx \frac{\tilde{v}}{g}.{\mathrm ''}
\label{eq:20}
\end{equation}

Here I shall not follow Carter's suggestion for estimating the value
of $\tilde{v}$ to use in deriving $\tilde{\ell}$ but shall use
instead the Press-Lightman suggestion $P \sim mv^3/h$.  Then the
connection between my $v$ and my $h$ will not be Carter's idea about
the pumping of blood (valid as that might be if one could get a
reasonable $\tilde{v}$), but rather the idea that the (land)
organism is assumed to be able to run and hence attain a velocity
$v$ that at least equals the speed of a pendulum of length $h$ (now
representing the leg length of the animal, which will be assumed to
be of the order of $h$) of large-amplitude swinging, so $v
\stackrel{>}{\sim} \sqrt{gh}$ and $P \stackrel{>}{\sim} m g^{3/2}
h^{1/2}$.

To put it another way, it is assumed that a running animal has
enough power to jump upward by an amount at least comparable to its
height $h$.  The energy for this is $E \sim m g h$, and the time $t$
over which this energy must be exerted must be less than the time to
fall a distance $h$, giving $t \stackrel{<}{\sim} \sqrt{h/g}$, so
the power must be $P \sim E/t \stackrel{>}{\sim} m g^{3/2} h^{1/2}$.

For the tallest running animal (e.g., a giraffe, or a human under
the approximation that both are large land animals of the same order
of magnitude of size), let us say that these inequalities for
running are saturated and that the resulting maximum value for $h$
is henceforth called $H$ (e.g., to avoid confusing it with the
estimate Press and Lightman \cite{PL} quote for $h$ from Press
\cite{Press}):
\begin{equation}
v \sim \sqrt{gH},
\label{eq:21}
\end{equation}
\begin{equation}
P \sim m g^{3/2} H^{1/2} \sim \alpha^3 m_e^3 m_p g^{3/2} H^{7/2},
\label{eq:22}
\end{equation}
where I have used Eq.\ (\ref{eq:19}) for the tallest running organism
mass $m$ in terms of its size $h=H$.

Now equating this estimate of the power necessary for running with
the estimated limit on the power from Eq.\ (\ref{eq:17}) for the
cooling rate, and also inserting the estimate of Eq.\ (\ref{eq:12})
for the acceleration of gravity $g$ on a habitable planet from
Press's requirements (i) and (ii) but not (iii), one gets that the
following estimate for the height of a giraffe:
\begin{equation}
H \sim \alpha^{4/5} m_e^{2/5} m_p^{-4/5} g^{-3/5}
\sim \alpha^{-7/10} m_e^{-19/20} m_p^{-13/20}
\sim \beta^{1/20} \gamma^{3/10} a_0
\approx 2.44\, {\mathrm m}.
\label{eq:23}
\end{equation}
Unlike the previous estimates of both Press \cite{Press} and Carter
\cite{Carter}, this estimate is within a factor of 2--3 of the height
of the tallest running animal, the tallest land animal, the giraffe.

It is interesting that when one writes the result as the Bohr radius
$a_0$ multiplied by the appropriate powers of $\alpha$, $\beta$, and
$\gamma$, the fine structure constant $\alpha$ drops out, and the
multiple of the Bohr radius involves only the ratio of electron and
proton masses, $\beta \equiv m_e/m_p$ and the ratio of the
electrical to gravitational forces between two protons, $\gamma
\equiv e^2/(Gm_p^2)$ (in units with $4\pi\epsilon_0 = 1$), with
Planck's constant not appearing anywhere beyond its appearance in
the Bohr radius $a_0 = 4\pi\epsilon_0\hbar^2/(m_e e^2)$.

It is also interesting that the power of the mass ratio $\beta$ is
so small.  Since $\beta^{1/20} \approx 0.6868$ is within a factor of
2 of unity, and since I have been cavalierly dropping many other
factors of 2, one can drop this factor in Eq.\ (\ref{eq:23}) to
obtain a simplified equation for the height of a giraffe that
actually works even better empirically (though it is still not quite
the maximum observed height of giraffes):
\begin{equation}
H \sim \gamma^{0.3} a_0
\approx 3.56\, {\mathrm m}.
\label{eq:24}
\end{equation}
That is, the height of a giraffe is here estimated to be roughly the
Bohr radius multiplied by the 0.3 power of the ratio of the
electrical and gravitational forces between two protons.

One can also use these estimates of the height of a giraffe to
estimate that the total number of nucleons (or atoms, which will be
of the same order, modulo the average atomic number of the molecules
that is yet another number of order unity that I am ignoring) in a
giraffe is either
\begin{equation}
N \sim \beta^{3/20} \gamma^{9/10} \approx 9.84\times 10^{31},
\label{eq:25}
\end{equation}
or, using the simplified Eq.\ (\ref{eq:24}) that drops the $\beta$
factor,
\begin{equation}
N \sim \gamma^{0.9} \approx 3.04\times 10^{32}.
\label{eq:26}
\end{equation}

The analogous masses are then
\begin{equation}
m \sim \beta^{3/20} \gamma^{9/10} m_p
\approx 165\,000 {\mathrm kg}
\label{eq:27}
\end{equation}
or
\begin{equation}
m \sim \gamma^{0.9} m_p
\approx 508\,000 {\mathrm kg}.
\label{eq:28}
\end{equation}

The first of these mass estimates, from the slightly more
sophisticated estimate of the height $H$, is reasonably good for
blue whales, but it is not supposed to apply to sea creatures but
only to running land creatures.  If one goes to elephants as the
most massive running land animal alive today, running up to 12\,000
kg, the mass estimates above are a bit too high, from about 14 to
about 42 times the largest recorded mass of a land animal species
alive today.  However, it is gratifying that even the mass estimates
are within two orders of magnitude of observed values.

From the first of my giraffe height estimates, Eq.\ (\ref{eq:23}),
one can also obtain estimates for the time to take a stride, the
stride time
\begin{equation}
t \sim \sqrt{H/g}
\sim \alpha^{2/5} m_e^{1/5} m_p^{-2/5} g^{-4/5}
\sim \alpha^{-8/5} m_e^{-8/5} m_p^{-1/5}
\sim \alpha^{-1} \beta^{-3/5} \gamma^{2/5} a_0/c
\approx 0.601 {\mathrm s}.
\label{eq:29}
\end{equation}
Carter \cite{Carter} has noticed that the characteristic time of our
mental and other biological processes is in Planck units of the
order of $m_e^{-1} m_p^{-1} \approx 0.0168$ seconds, which
coincidentally is only 0.57\% larger than the period of the 60-Hertz
alternating current used in North America, since $m_e m_p \approx
59.66$ Hz.  One may express the stride time above in terms of
Carter's simple characteristic time as
\begin{equation}
t \sim \alpha^{-3/2} \beta^{-3/5} \gamma^{-1/10} m_e^{-1} m_p^{-1}
\approx 35.8\, m_e^{-1} m_p^{-1}.
\label{eq:30}
\end{equation}

The stride time thus has a slightly less positive power of the
gravitational coupling (in $\gamma^{-1}$) than Carter's
characteristic time (which by itself would tend to make the stride
time shorter than Carter's time), but the difference in the powers
of the fine structure constant $\alpha$ and the electron-proton mass
ratio $\beta$ gives a large factor, nearly $1.46\times 10^5$, whose
logarithm is about 1.43 times larger than the negative of the
logarithm of the $\gamma^{-1/10} \approx 0.000\,246$, so in the end
the stride time estimate comes out nearly 40 times larger than
Carter's electrifyingly simple characteristic time.

One can similarly calculate the running velocity
\begin{eqnarray}
v &\sim& H/t \sim \sqrt{Hg}
\sim \alpha^{2/5} m_e^{1/5} m_p^{-2/5} g^{1/5}
\sim \alpha^{-9/10} m_e^{-13/20} m_p^{-9/20} \nonumber \\
&\sim& \alpha \beta^{13/20} \gamma^{-1/10} {\mathrm c}
\approx 1.36\times 10^{-8} {\mathrm c}
\approx 4.07\, {\mathrm m/s} \approx 14.6\, {\mathrm km/hr}
\approx 9.10\, {\mathrm mph} .
\label{eq:31}
\end{eqnarray}
This purely theoretically derived estimate for the fastest animal
running velocity is perhaps roughly the speed that a theorist like
me can run, but it is slightly less than 40\% of the average speed
of 10.35 m/s or 37.27 km/hr or 23.16 mph (miles per hour) of Michael
Johnson in running 200 meters in 19.32 seconds, and it is less than
20\% of the maximum speed of a cheetah.

It is interesting that if one compares this estimate for the top
running speed of an animal with the remarkably good estimate of Eq.\
(\ref{eq:11}) for the speed of a low satellite, one finds that
\begin{equation}
v \sim (\beta\gamma)^{-1/10}\, v_{\mathrm satellite} 
= \beta^{-1/5}
\left(\frac{G m_e m_p}{e^2}\right)^{0.1}\, v_{\mathrm satellite},
\label{eq:32}
\end{equation}
so the speed of a running animal is estimated to be just a few times
larger (by a factor of $\beta^{-1/5} \approx 4.5$) than the speed of
a low satellite multiplied by the one-tenth power of the ratio of
the gravitational to the electrical attractions between the electron
and the proton in a hydrogen atom.  It can perhaps give one a bit of
a feel for this tiny ratio of forces if one realizes that it is even
millions of times smaller (by a factor of $\beta^2 \approx 0.3\times
10^{-6}$) than the tenth power of the ratio of the speed that one can
run to the speed of a satellite.

\section{Using Anthropic Estimates for the Charge and Mass of the
Electron and Proton}

So far I have used the observed values of the charge and mass of the
electron and proton to give new estimates for the height, stride
time, and running velocity of the giraffe, the tallest running
animal.  However, one can also derive anthropic estimates for all of
these quantities \cite{Page} that do not depend upon {\it any}
measured continuous parameters or coupling constants, so it may be of
interest to insert these purely mathematical values into the
estimates above.

Basically, \cite{Page} uses the anthropic results of
\cite{Carter1,Carter2,Carr-Rees} and the renormalization group
formulas of \cite{MarCen} to derive, using no fudge factors at all,
\begin{equation}
\alpha\ln{\alpha} \sim -{\pi\over 100},
\label{eq:33}
\end{equation}
\begin{equation}
\beta \sim \alpha^2,
\label{eq:34}
\end{equation}
\begin{equation}
\gamma \sim \alpha^{-19},
\label{eq:35}
\end{equation}
\begin{equation}
m_e \sim \alpha^{12},
\label{eq:36}
\end{equation}
\begin{equation}
m_p \sim \alpha^{10},
\label{eq:37}
\end{equation}
\begin{equation}
a_0 \sim \alpha^{-13},
\label{eq:38}
\end{equation}
\begin{equation}
M_s = L_s \sim \alpha^{-20},
\label{eq:39}
\end{equation}

Let us take the solution of Eq.\ (\ref{eq:33}), with an equal sign
rather than the $\sim$ sign, as $\alpha_a$ (with subscript $a$ for
``anthropic''), with the numerical solution
\begin{equation}
\alpha_a \approx 0.006\,175\,533\,381.
\label{eq:40}
\end{equation}

This then gives
\begin{equation}
\alpha \sim \alpha_a \approx 0.846 \alpha,
\label{eq:41}
\end{equation}
\begin{equation}
1/\alpha \sim 1/\alpha_a \approx 162 \approx 1.18/\alpha,
\label{eq:42}
\end{equation}
\begin{equation}
\beta \sim \alpha_a^2 \approx 0.000\,038 \approx 0.070\, \beta,
\label{eq:43}
\end{equation}
\begin{equation}
\gamma \sim \alpha_a^{-19} \approx 9.5\times 10^{41}
\approx 770\,000\, \gamma,
\label{eq:44}
\end{equation}

Then my estimate for the giraffe height would be
\begin{equation}
H \sim \alpha_a^{-5.6} a_0 \approx 2.4\times 10^{12} a_0
\approx 125\, {\mathrm m},
\label{eq:45}
\end{equation}
where for the last number I am assuming that if $\alpha$, $\beta$,
and $\gamma$ were changed to the `anthropic' values given here, then
the meter would be defined so that it still equaled $1.89\times
10^{10}\, a_0$.  This would then imply that the purely theoretical
anthropic estimate for the number of atoms in the largest land animal
in a typical biophilic region of the universe, multiverse, or
holocosm would be of the order of
\begin{equation}
N \sim \alpha_a^{-16.8} \sim 10^{37}.
\label{eq:46}
\end{equation}

A number of this rough order of magnitude might be a crude estimate
for the maximally complex living being in the holocosm.  It seems
plausible that there might be significantly more complex beings that
that, but also that their numbers might be tailing off sufficiently
rapidly that this number gives a rough upper limit.

\section*{Acknowledgments}

	I appreciated the hospitality of Edgar Gunzig and the
Cosmology and General Relativity symposium of the Peyresq Foyer
d'Humanisme in Peyresq, France, where I had useful discussions on
these issues with Brandon Carter.  I also am grateful for email
discussions with Carter and for his sending me a copy of his recent
paper \cite{Carter}.  This research was supported in part by the
Natural Sciences and Engineering Research Council of Canada.


\begin{thebibliography}{99}

\bibitem{Press} W.~H.~Press, Am.\ J.\ Phys.\ {\bf 48}, 597 (1980).

\bibitem{PL} W.~H.~Press and A.~P.~Lightman, Phil.\ Trans.\ R.\ Soc.\
A {\bf 310}, 323 (1983).

\bibitem{Carter} B.~Carter, ``Objective time and subjectivity
temporality in anthropic reasoning,'' Contribution to {\em Time in
Science, Anthropology, Religion, Arts}, SR21 Workshop, Thessaloniki
and Athenes.

\bibitem{BT} J.~D.~Barrow and F.~J.~Tipler, {\em The Anthropic
Cosmological Principle} (Oxford University Press, New York, 1986),
pp.\ $19^2$f.

\bibitem{Page} D.~N.~Page, ``Anthropic estimates of the charge and
mass of the proton,'' arXiv:hep-th/0302051.

\bibitem{Carter1}  B.~Carter, in M.~S.~Longair (ed.), {\em
Confrontation of Cosmological Theory with Observational Data}
(Riedel, Dordrecht, 1974), pp.\ 291-298.

\bibitem{Carter2}  B.~Carter,
``Large numbers in astrophysics and cosmology''
(paper presented at Clifford Centennial Meeting, Princeton, 1970).

\bibitem{Carr-Rees}  B.~J.~Carr and M.~J.~Rees,
Nature {\bf 278}, 605-612 (1979).

\bibitem{MarCen}  W.~J.~Marciano and G.~Senjanovi\'c,
Phys.\ Rev.\ {\bf D25}, 3092-3095 (1982).

\end{thebibliography}
\end{document}